\documentstyle[11pt,paspconf,epsfig,amssymb]{article}

\begin{document}
\title{Unifying All Blazars}

\author{Giovanni Fossati\altaffilmark{1}}
\affil{International School for Advanced Studies (SISSA/ISAS), 
via Beirut 4, 34014 Trieste, Italy }
\altaffiltext{1}{moving to Center for Astrophysics and Space Sciences
(CASS), University of California, San Diego, 9500 Gilman Dr., 
La Jolla CA 92093-0424, USA }


\begin{abstract}
We propose a new physical model for the unification of all Blazars. 
The blazar phenomenology can be accounted for by a sequence in the source
power and intensity of the diffuse radiation field surrounding the
relativistic jet. 
This regulates the equilibrium electron distribution and hence the shape of
the spectral energy distribution (SED), and in turn the classification
into a blazar ``flavour". 
The processes inside and beside the emitting jet are the same throughout
the blazar class. 
Objects along this sequence would be observationally classified as
High-frequency-peaked BL Lac objects (HBL), Low-frequency BL Lac objects
(LBL), Highly Polarized Quasars (HPQ) and Lowly Polarized Quasars (LPQ).
\end{abstract}

\section{Introduction}

Among AGNs blazars represent the most extreme and powerful sources. 
The fundamental property characterizing blazars is their beamed continuum,
due to plasma moving relativistically along the line of sight.
This scenario seems to apply to objects with somewhat different
observational properties leading to different classifications/definitions. 
Objects with significant emission line equivalent widths are usually found
as flat spectrum radio quasars (FSRQ). 
Objects without emission lines (EW $<$ 5 \AA) are classified as BL Lac objects.
Different flavours of BL Lac objects have been found in radio and
X-ray surveys. 
These also correspond to differences in the overall spectral energy
distributions (SED) leading to the sub-classification of High and Low
frequency peaked BL Lacs (e.g. Padovani \& Giommi 1995).
Nevertheless, while different sub--classes have different average
properties, the actual distinction among them is certainly fuzzy and
so far several sources have shown intermediate behaviour. 
In fact arguments for a substantial ``continuity" in the continuum spectral
properties leading to adopt the blazar denomination as including both, BL
Lacs as well as FSRQs, have been recently re--proposed
(e.g. Sambruna et al. 1996, Fossati et al. 1998).

The recent discovery of about $\sim$60 blazars emitting in the
$\gamma$--ray band (GeV and TeV, by CGRO and Cherenkov telescopes), 
have revealed that the bulk of their radiative output is emitted in the
$\gamma$--ray range, thus allowing us to discuss for the first time the
characteristics of blazars knowing their total emission output and their
entire SED. 

\noindent
\underbar{\bf The goal:}
we wanted to try to derive general properties of the continuum and
understand whether and how they differ for instance in BL Lac objects and
FSRQ. 
We consider blazars as a single class/family of objects, assuming
that the same physical mechanisms operate in relativistic jets over a wide
range of luminosities.
We addressed these issues from two sides: 

\settowidth{\labelwidth}{$\rhd$}
\hangindent \labelwidth
\noindent
\fbox{A}$\rhd$ a purely observational approach based on complete
sub--samples of blazars aimed at studying the systematics of
the SEDs of blazars from radio to $\gamma$-rays. 

\hangindent \labelwidth
\noindent
\fbox{B}$\rhd$ a theoretical approach based on modelling
individually the SEDs of the $\gamma$--ray sources.  
The goal is not to model spe\-ci\-fic bla\-zars, but to unveil
possible trends among physical quantities, sheding light on the
relationship among different sub--classes.

\noindent
Here we present a summary of a few highlights of this work, referring 
to Fossati et al. (1998), and Ghisellini et al. (1998) for the full discussion.

\noindent
\underbar{\bf The samples and the Data: }
we considered a ``total blazar sample" resulting from the merging
of the following three samples: the Slew Survey sample and the 
1 Jy sample of BL Lac objects and the FSRQ sample derived from the 2 Jy 
radio sample of Wall \& Peacock (1985).
We consider objects irrespective of their original classification 
The net number of sources is 126.

\begin{figure}
\centerline{
\epsfig{figure=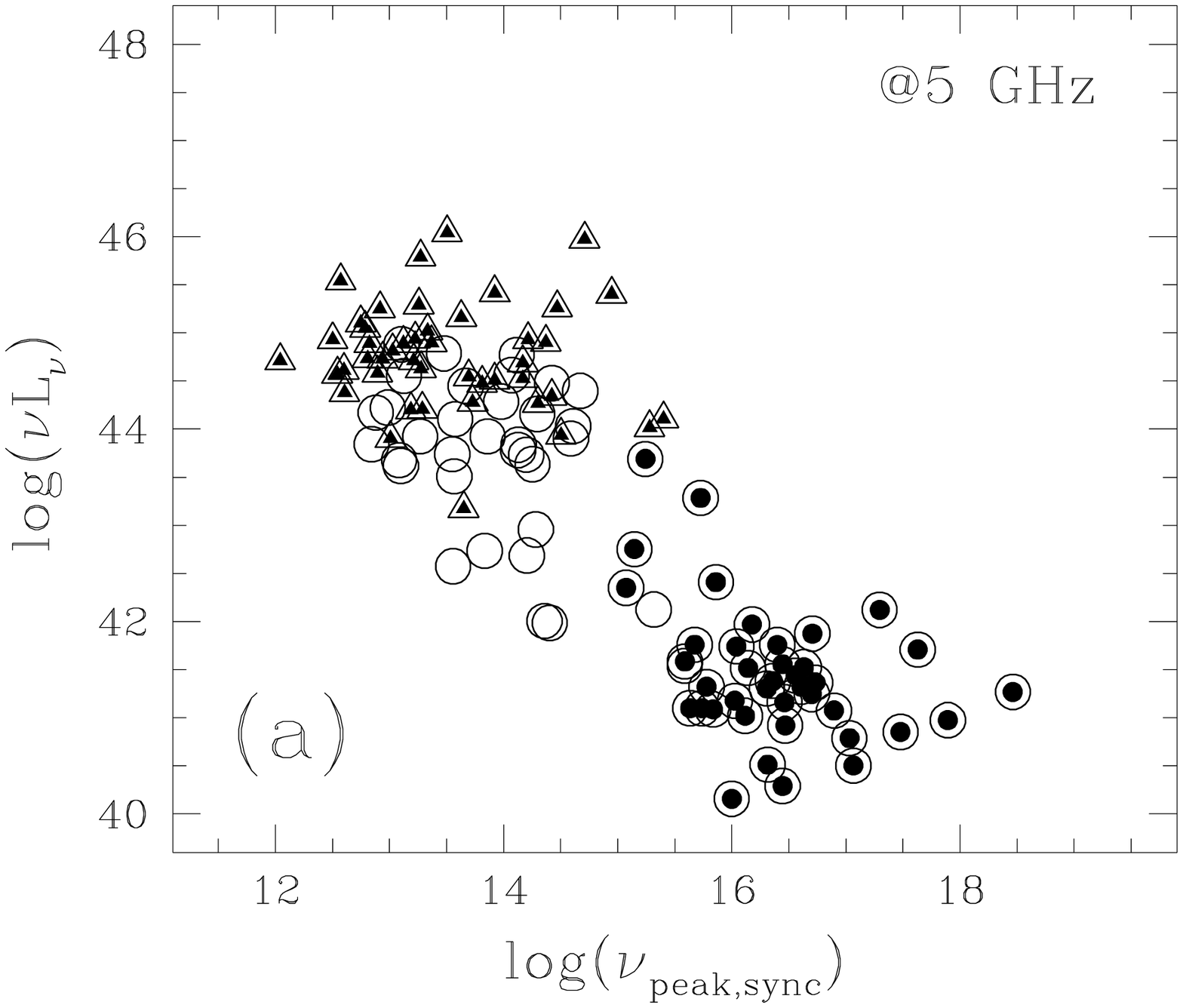,width=0.48\linewidth}\hfill
\epsfig{file=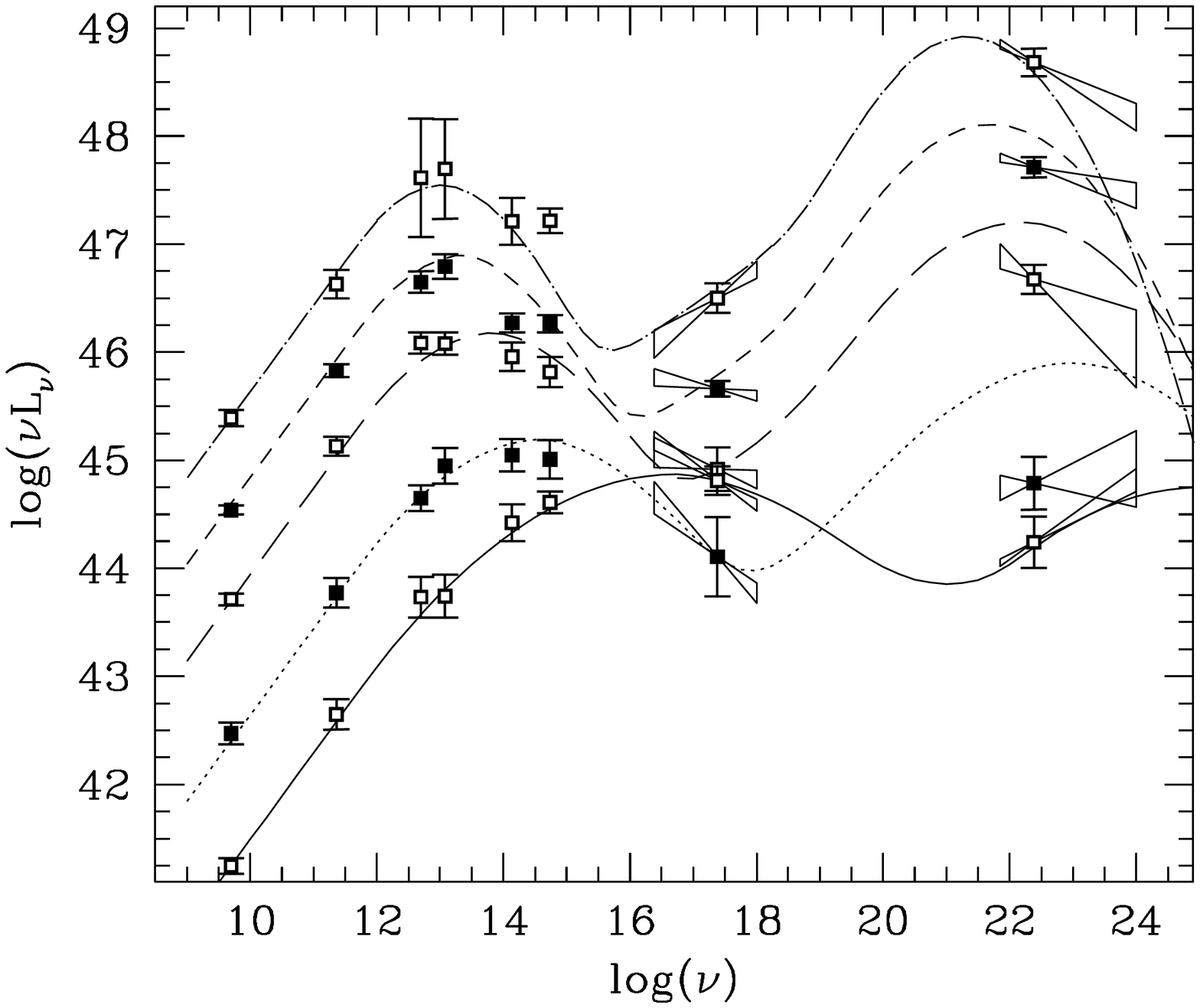,width=0.48\linewidth}
}
\vspace{-1.7cm}
\begin{minipage}[t]{0.52\linewidth}
\caption{\footnotesize\sf\protect\baselineskip 8pt 
The peak frequency of the synchrotron component
plotted against radio luminosity.}
\label{fig:peak_lr}
\end{minipage} \hfill
\begin{minipage}[t]{0.52\linewidth}
\caption{\footnotesize\sf\protect\baselineskip 8pt 
Average SEDs.
Lines are re\-pre\-senta\-ti\-ves
of the one-parameter-family discussed in text. }
\label{fig:sed_medie}
\end{minipage}
\end{figure}

\section{Results \fbox{A}} 

\noindent
\underbar{\bf $\nu_{\rm peak}$ vs. Luminosity}
We estimated the position of the peak of the synchrotron
component in individual objects by fitting the SED of each source 
($\nu$ vs. $\nu$L$_\nu$) with a third degree polynomial.
The resulting peak frequencies are plotted in Fig.~\ref{fig:peak_lr}
versus the radio luminosity.  
Strong correlation (P$\gg$ 99.99 \%) is present in the sense of $\nu_{\rm
peak,sync}$ decreasing with increasing luminosity. 
The same trend holds also for $\gamma$--ray luminosity, for the
restricted sample of $\gamma$--ray detected sources.
Since on one hand in flux limited samples spurious correlations can be
introduced by the luminosity/redshift relation and on the other hand the
correlations might be due to evolutionary effects genuinely related to
redshift, we also restricted the analysis to sources in the $z < 0.5$
interval, and applying the partial correlation algorithm.
In both cases the correlation still holds.

\noindent
\underbar{\bf Average SEDs}
Since luminosity appears to have an important role in that it correlates
with the main  spectral parameters we binned the ``total blazar
sample" according to 5~GHz radio luminosity.
It may be desirable to use the {\it bolometric} luminosity which in all
cases is close to the {\it $\gamma$--ray} one, available for only a few
objects. 
This latter correlates with the {\it radio} luminosity (although this is a
debated issue, e.g. Fossati et al. 1998), which could then represent a
suitable approximation. 
The resulting SEDs are shown in Fig.~\ref{fig:sed_medie}, where
we superimposed ``analytic" SEDs which are representatives of a family of
curves parameterized only on the radio luminosity, which in
turn possibly traces the bolometric one.
The ``core" of this parameterization is the assumption that the peak
\textit{frequency of the synchrotron spectral component is (inversely)
related to radio luminosity} (see Fig.~\ref{fig:peak_lr}).
The high energy spectral component has been "placed" with
respect to the radio--to--optical one assuming that:  
(a) the ratio of the frequencies of the high and low energy peaks is constant 
($\simeq 5\times10^8$), 
(b) high $\gamma$--ray peak and radio luminosities have a fixed ratio, 
$\simeq 3\times10^3$. 
Given the extreme simplicity of the assumptions, it is remarkable
that the phenomenological model describes reasonably well the average SEDs.

\noindent
\fbox{
\parbox[t]{0.98\textwidth}{
\underbar{\bf Highlights}

\hangindent \labelwidth
$\rhd$ blazar SEDs have two peaks, whose frequency
ratio is compatible with being constant.  
The lower energy one  shows an anti-correlation with luminosity.

\hangindent \labelwidth
$\rhd$ the $\gamma$--ray dominance L$_{\rm Comp}$/ L$_{\rm sync}$ 
increases with increasing power.

\hangindent \labelwidth
$\rhd$ blazar continua can be described as a \textit{one parameter family} 
of curves with luminosity as the fundamental parameter.

\noindent
These trends suggest
that we deal with a {\it continuous spectral sequence} within the 
blazar family, rather than with separate spectral classes.
}}

\section{Results \fbox{B}}

Strong correlations have been found among the physical parameters
derived from the External Radiation Compton model (ERC).
The most interesting quantity to investigate is the Lorentz factor at the
break of the electron distribution, $\gamma_{\rm peak}$.
It determines the location of both the synchrotron and the
Compton peaks, and therefore largely determines the shape of the SED. 
Two are the results of particular interest:

\hangindent \labelwidth
\noindent
$\rhd$ $\gamma_{\rm peak}$ correlates with the total
energy density in the emitting region: 
$\gamma_{\rm peak}\propto$ (U$_{\rm r}$+U$_{\rm B})^{-0.6}$.
One way to explain this is to assume that $\gamma_{\rm peak}$ is the
result of a competition between the radiative cooling and the
(re--)acceleration process. 
The typical emitting electron would be quickly accelerated up
to the energy where cooling is important, while only a few particles would
be accelerated at higher energies. 

\hangindent \labelwidth
\noindent
$\rhd$ $\gamma_{\rm peak}$ correlates with the Compton dominance. 
This link can be interpreted as the consequence of a change in the
radiation energy density of the external field. 
An increase in the latter in fact leads to an increase in the particle
Compton cooling and therefore both to a decrease in $\gamma_{\rm peak}$
and a relative increase in the $\gamma$--ray luminosity. 

\hangindent \labelwidth
\noindent
$\rhd$ different sub--classes of blazars are located in different areas 
of this correlation.  
HBLs at the extreme and LBLs smoothly overlapping with FSRQ.

\noindent
The possibility that the external photon field involved in the
ERC process is related to the radiation reprocessed as broad
emission lines, seems to be at least qualitatively in agreement with
the observational evidence concerning the emission line luminosity in
the suggested blazar sequence.

\section{The Blazar Unification}

\begin{minipage}{0.48\linewidth}
The main conclusion is that despite the phenomenological differences among
different sub--classes of blazars, a unitary scheme is possible, with the
evidence for a well defined sequence in the properties of HBL, LBL and
FSRQ with increasing importance of an external radiation field. 
The correlations among the different quantities ensure that the
knowledge of one of them allows to estimate the entire spectral energy
distribution, and also the probable classification of the object.
The \textit{proposed blazar unifying sequence} can be therefore
summarized as follows: \\
\end{minipage}
\hfill
\begin{minipage}{0.48\linewidth}
\epsfig{file=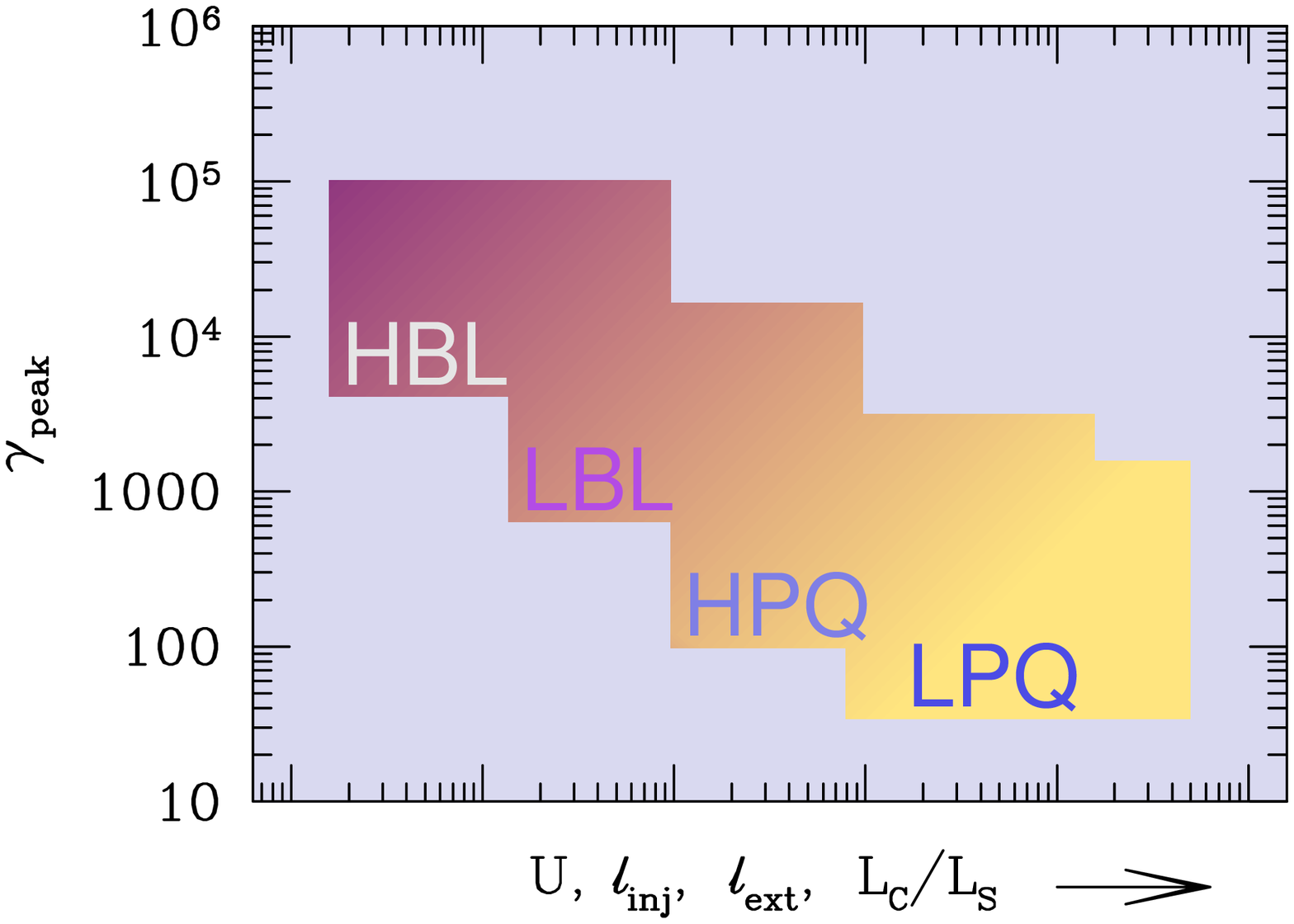,width=\linewidth}  
Figure 3. 
\footnotesize\sf
Schematic representation of the proposed unifying scheme. 
\vfill
\end{minipage}

\centerline{
\fbox{
\begin{minipage}[t]{0.29\linewidth}
\sloppy\footnotesize
\fbox{HBL:} are the sources with the lo\-west in\-trin\-sic po\-wer 
and the weakest exter\-nal radiation field (no or weak emission lines).  
Cooling is less dramatic and electrons can reach energies
high enough to produce soft X--ray synchrotron emission and TeV radiation
through the IC process. 
Being the inverse Compton cooling ineffective, the Compton dominance is
expected to be small.
\end{minipage}
}
\hfil
\fbox{
\begin{minipage}[t]{0.29\linewidth}
\sloppy\footnotesize
\fbox{LBL:} are intrinsically more powerful than HBL. 
The external field can be responsible for most of the cooling. 
The stron\-ger cooling limits electrons energy implying that the
synchrotron and inverse Compton emission peak at lower frequencies, 
in the optical and GeV bands, respectively, with a larger Compton
dominance.
\end{minipage}
}
\hfil
\fbox{
\begin{minipage}[t]{0.29\linewidth}
\sloppy\footnotesize
\fbox{FSRQ} are the most powerful blazars. 
The contribution from the external radiation to the cooling is the
gre\-a\-test. 
Syn\-ch\-ro\-tron and IC emission cannot extend at fre\-quen\-cies lar\-ger
than the IR and MeV--GeV bands. 
$\gamma$--ray radiation dominates the ra\-dia\-ti\-ve out\-put. 
Within this class, there is the hint for LPQ to be more extreme. 
\end{minipage}
}
}

\end{document}